\documentclass[pra,twocolumn,showpacs,preprintnumbers,amsmath,amssymb]{revtex4}
\usepackage{amsfonts}
\usepackage{graphicx}
\usepackage{dcolumn}
\usepackage{float}
\usepackage{bm}

\renewcommand{\theequation}{{\arabic{section}.\arabic{equation}}}
\makeatletter\@addtoreset{equation}{section} \makeatother

\newcommand{\ind}[1]{\textrm{#1}}
\newcommand{\bra}[1]{{\langle #1 |}}
\newcommand{\ket}[1]{{| #1 \rangle}}

\newcommand{\com}[1]{}

\renewcommand{\O}[0]{\mathcal{O}}

\begin{document}

\title{Quantum Nucleation and Macroscopic Quantum Tunneling in
Cold-Atom Boson-Fermion Mixtures}

\author{Dmitry Solenov\footnote{E-mail: solenov@lanl.gov}
and Dmitry Mozyrsky\footnote{E-mail: mozyrsky@lanl.gov}}
\affiliation{Theoretical Division (T-4), Los Alamos National
Laboratory, Los Alamos, NM 87545, USA}

%\date{\today}

\begin{abstract}
Kinetics of phase separation transition in boson-fermion cold atom
mixtures is investigated. We identify the parameters at which the
transition is governed by quantum nucleation mechanism,
responsible for the formation of critical nuclei of a stable
phase. We demonstrate that for low fermion-boson mass ratio the
density dependence of quantum nucleation transition rate is
experimentally observable. The crossover to macroscopic quantum
tunneling regime is analyzed. Based on a microscopic description
of interacting cold atom boson-fermion mixtures we derive an
effective action for the critical droplet and obtain an asymptotic
expression for the nucleation rate in the vicinity of the phase
transition and near the spinodal instability of the mixed phase.
We show that dissipation due to excitations in fermion subsystem
play a dominant role close to the transition point.
\end{abstract}


%05.30.Jp Boson systems (for static and dynamic properties of
%Bose-Einstein condensates, see 03.75.Hh and 03.75.Kk)

%03.75.Kk Dynamic properties of condensates; collective and
%hydrodynamic excitations, superfluid flow

%03.75.Mn Multicomponent condensates; spinor condensates

%[no] 32.80.Pj Optical cooling of atoms; trapping
%37.10.Gh Atom traps and guides

%67.90.+z Other topics in quantum fluids and solids; liquid and
%solid helium (restricted to new topics in section 67)

%87.80.Cc Optical trapping

\maketitle

\section{Introduction}\label{sec:intro}

Macroscopic metastable states of trapped cold atom systems have
been a subject of active experimental and theoretical study for
more than a decade \cite{UedaLeggett}. Unlike a homogeneous system
of bosons, where infinitesimally small attractive interaction
between atoms leads to a collapse, trapped bosons are known to
form long lived Bose-Einstein Condensates
\cite{UedaLeggett,Modugno} (BEC) due to zero-point energy which,
for sufficiently low densities, can compensate the negative
interaction energy thus maintaining the system in equilibrium.
Upon increasing the BEC density, interaction energy grows, and, at
some instability point (i.e., at a certain number of particles in
the trap $N_c$, with $N_c\sim 10^3$ for a typical trap),
zero-point energy can no longer sustain the negative pressure due
to the interactions and the system collapses. It has been argued
in the literature \cite{UedaLeggett} that near the instability
point (for BEC densities slightly lower than the instability
density), the effective energy barrier that prevents BEC from
collapsing becomes so low that the system can quantum mechanically
tunnel into the dense (collapsed) state.

Such phenomenon of Macroscopic Quantum Tunneling (MQT), however,
has never been observed experimentally due to a strong dependence
of the barrier height on the total number of particles in the trap
($N$). Indeed it has been shown \cite{UedaLeggett} that the
tunneling exponent for such a transition near the instability
point scales as $N(1-N/N_c)^{5/4}$ and therefore very fine tuning
of the total particle number $N$ is required in order to keep the
tunneling exponent relatively small [$(1-N/N_c)\ll 1$]. Since for
most BEC setups the total number of the trapped atoms fluctuates
and typically obeys Poissonian statistics, the error in $N$ scales
as $N^{1/2}$ and therefore such a stringent requirement is hard to
fulfill. Thus the system is typically either in a sub-critical
state with no barrier present ($N\ge N_c$) or is in the state with
very high energy barrier and therefore very low MQT rate.

In this paper we propose another paradigm for observation of
tunneling driven phase transition effects in cold atom systems
based on the theory of quantum nucleation
\cite{LifshitzKagan,LifshitzKhokhlov,Burmistrov}. It has long been
known that a mixture of $^3$He-$^4$He undergoes a phase separation
transition at relative concentration of $^3$He in $^4$He of around
$6\%$ at temperatures close to the absolute zero~\cite{Edwards}.
Since such a phase separation is a first order phase transition
(it is observed to be accompanied by the latent heat release down
to mK temperatures), the order parameter must have some finite
(microscopic) correlation length and therefore the transition is
expected to occur through the formation of nuclei of the new
stable phase in the old metastable one. As usual, dynamics of the
nucleation process is controlled by the competition of the surface
and bulk energies of the nuclei and therefore, in order for a
given nucleus to become stable (supercritical), it must overcome a
potential barrier formed by the two above contributions. While in
most systems such a transition is a thermally activated process,
it has been argued that in the $^3$He-$^4$He mixture at
sufficiently low temperatures (below 100 mK) the transition is
driven by the quantum tunneling. In particular, it was predicted
\cite{LifshitzKagan,Burmistrov} that near the transition line the
tunneling exponent for such a transition rate is proportional to
$\Delta\mu^{-7/2}$, where $\Delta\mu $ is the difference in the
chemical potentials of the two phases. It has later been found
experimentally \cite{HeExperiment} that below 80 mK kinetics of
such phase separation transition becomes independent on
temperature and therefore it must be driven by the quantum
tunneling. However, the experiments have been unable to verify the
expected dependence of the nucleation rate on the systems's
parameters (i.e., $\Delta\mu$, etc) - partly due to the poor
knowledge of microscopic interactions between particles in such a
strongly correlated system.

We argue that contemporary cold atom systems provide an excellent
candidate for studying and observing the kinetics of such a phase
separation transition in boson-fermion mixtures \cite{SELF}.
Mixtures of boson and fermion atoms are typically realized in
experiments studying fermionic superfluidity, where bosons play
role of a coolant \cite{BosCool}. Another interesting realization
of boson-fermion mixture has been demonstrated in two-component
fermion system, where strongly bound Cooper pairs correspond to
bosons interacting with unpaired fermion atoms \cite{Ketterle}. In
the present paper we begin with detailed derivation and analysis
of the results outlined in Ref.~\onlinecite{SELF}. Significant
attention is given to supercritical dynamics, which reflects the
dissipative mechanisms and is measurable in less interacting
system.

Starting from a microscopic description of a boson-fermion mixture
we derive an effective action for the order parameter (the BEC
density) taking into account fermion-boson interaction. We show
explicitly that the classical potential for the order parameter
due to such interaction has two minima corresponding to the two
phases of the system (mixed and phase separated), see
Sec.~\ref{sec:phases}. We analyze the coherence length associated
with the system and demonstrate that it varies from finite to
divergent and therefore allows different mechanisms for the phase
transition, see Sec.~\ref{sec:equil}. At low fermion densities the
two minima of the potential are separated by the finite energy
barrier resulting in finite coherence length, which points out
that such a transition is of the first order
\cite{bosons,Timmermans}. We then derive an expression for the
nucleation (tunneling) rate of the critical droplet of the pure
fermion phase near the phase transition line and near the line of
absolute (spinodal) instability  of the mixed phase in
Secs.~\ref{sec:i-ii} and \ref{sec:spinodal}. We show that the
transition rate is measurable (of the order
$0.1-1$s$^{-1}\mu$m$^{-3}$) for densities reasonably close to the
phase transition line and small fermion-boson mass ratio. The
limitation on the mass ratio comes as the result of dissipative
dynamics due excitations in fermion system. Near the phase
transition line it leads to significant modification (increase) of
the transition rate for the system with high fermion-boson mass
ratio, in which case the observable transition rates exist closer
to the line of absolute instability (where the dissipation is less
effective and the tunneling exponent becomes reasonably small).
Our results for this regime are similar to those obtained in
Ref.~\onlinecite{UedaLeggett} for the MQT in the systems of
trapped bosons with attractive interactions. However, in the case
of boson-fermion separation transition the height of the potential
barrier and, thus, the tunneling exponent are controlled not by
the total number of particles in the trap, but their densities,
scaling as $(1-n/n_c)^{1/2}$. Therefore, for sufficiently large
numbers of particles in the trap the tunneling exponent can be
fine-tuned with a desired accuracy which makes it possible to
observe the MQT rate in a well controlled and predictable regime.

\section{Model}\label{sec:model}

We consider a Bose-Einstein Condensate (BEC) interacting with a
single species of fermions (in the same spin state). Interactions
in such the mixture are characterized by two scattering lengths
$a_{BB}$ and $a_{BF}$. Fermions and bosons interact through
contact potential $\lambda_{BF}\delta(\mathbf{r}-\mathbf{r}')$,
contributing term
$\lambda_{BF}\psi_B^\dag\psi_F^\dag\psi_B\psi_F$, where
$\lambda_{BF} = 2\pi\hbar^2a_{BF}(1/m_B + 1/m_F)$; $\psi_B$ and
$\psi_F$ are boson and fermion fields respectively. In addition,
boson-boson particle interaction give rise to another term
$\lambda_{BB}\psi_B^\dag\psi_B^\dag\psi_B\psi_B/2$ in the
Hamiltonian density, with $\lambda_{BB}=4\pi\hbar^2a_{BB}/m_B$.
The direct coupling between fermions is negligible: s-scattering
channel is forbidden for the fermions in the same spin state,
while p-wave scattering is small compared to boson-boson and
boson-fermion contact interaction. The potential part of the
energy density can be cast in the form
%/////////////////////////////////////////////////////
\begin{eqnarray}\label{eq:intro:Eraw}
E_p = &-& \mu_B|\psi_B|^2 -\mu_F\psi_F^\dag\psi_F
\\\nonumber
&+& \frac{\lambda_{BB}}{2}|\psi_B|^4 +
\lambda_{BF}|\psi_B|^2\psi_F^\dag\psi_F
\end{eqnarray}
%/////////////////////////////////////////////////////
For the purposes of the present calculation we can neglect the
spatial dependence of the trapping potential and assume that the
local densities of fermions and bosons are set by the constant
chemical potentials $\mu_F$ and $\mu_B$. Indeed, since the
nucleation occurs at finite coherence length (to be defined
below), the shape of the trapping potential should play little
role in the dynamics of the phase transition as long as the
effective size of the trap is much greater than the coherence
length.

It is convenient to describe the system in terms of the boson
field only, tracing $e^{-H/k_BT}$ ($H$ is the overall Hamiltonian)
with respect to the fermion field. Such averaging can be easily
carried out within mean field, i.e., the Thomas-Fermi
approximation \cite{Mahan,Viverit,Mozyrsky}, so that the
calculation reduces to the evaluation of the canonical partition
function (or free energy) of the free fermions with effective
chemical potential $\tilde\mu_F=\mu_F-\lambda_{BF}|\psi_B|^2$.
Later, in Sec.~\ref{sec:equil}, we will proceed beyond this
approximation and account for the fermion excitations interacting
with the condensate. For the purpose of this and the following
section such a correction is not necessary.

Within Thomas-Fermi approximation \cite{Mahan} we deal with the
energy density of free fermions in the zero-temperature limit. It
can be easily found differentiating fermion free energy $\sqrt
2Vm_F^{3/2}\tilde\mu_F^{5/2}/{5\pi^2\hbar^3}$ with respect to the
volume (note that $\tilde\mu_F\sim V^{-2/3}$). We finally obtain
the effective classical potential density for the bosons
%/////////////////////////////////////////////////////
\begin{eqnarray}\label{eq:intro:Eeff}
E(\rho) =  &-& \mu_B \rho  + \frac{1}{2}\lambda_{BB} \rho^2
\\ \nonumber
&-& \lambda_0 \left(\mu_F  - \lambda_{BF} \rho\right)^{5/2}
\theta(\mu_F - \lambda_{BF}\rho),
\end{eqnarray}
%/////////////////////////////////////////////////////
where $\theta(x\geq 0)=1$, $\theta(x<0)=0$, $\lambda_0 =
(2m_F)^{3/2}/15\pi^2\hbar^3$, and we have used the density--phase
variables, i.e. $\psi_B=\sqrt\rho e^{i\phi}$, for the condensate
field. The approximation discussed above ignores gradient terms in
the diagrammatic expansion of the partition function
\cite{Kleinert,ImambekovDemler}. This terms are not important for
the phase structure of the equilibrium system and will be
discussed in Sec.~\ref{sec:equil}.

In the next section we discuss the structure of the phase diagram
describing possible configurations in the equilibrium
boson-fermion system. Subsequent sections are devoted to the
kinetics of the transition. The transition rates are calculated in
Secs.~\ref{sec:i-ii} and \ref{sec:spinodal}. Sections
\ref{sec:thermal} and \ref{sec:expansion} are devoted to the
discussion of thermal activation and supercritical expansion. The
results are discussed in Sec.~\ref{sec:discus}.

\section{System in Equilibrium: The Phase Diagram}\label{sec:phases}

Equation (\ref{eq:intro:Eeff}) was obtained tracing over the
fermionic part of the partition function. In this respect, it can
be viewed as effective potential energy of boson condensate. The
structure of the fast fermionic part of the system, in particular
the density of fermions, is directly related to the state of the
boson condensate. In the limit of zero temperatures we obtain the
fermion density in the form
%////////////////////////////////////
\begin{equation}\label{eq:phases:nF-gen}
\rho_F(\rho) = \frac{5}{2}\lambda_0 (\mu_F-\lambda_{BF}\rho)^{3/2}
\theta (\mu_F-\lambda_{BF}\rho)
\end{equation}
%////////////////////////////////////
where the chemical potential $\mu_F$ is the same for all parts of
the system in the mixture as far as the time scale of slow boson
subsystem is of interest.

\begin{figure}
\includegraphics[width=8.1cm]{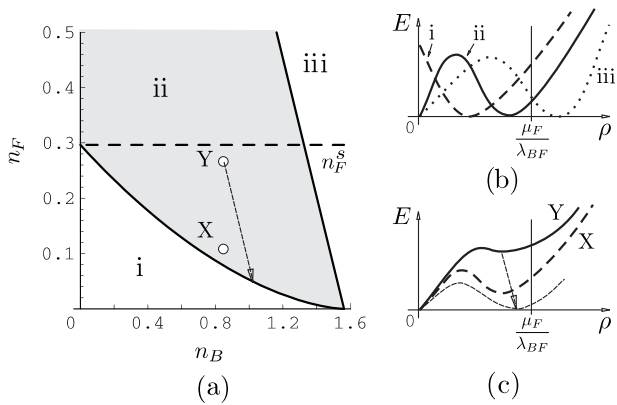}%[width=8.1cm]
\caption{The phase diagram of the uniform boson-fermion mixtures.
Three phase are possible: (i) mixture, (ii) coexistence of pure
fermions with mixed fraction, (iii) coexistence of pure fermion
and pure boson fractions. The arrow indicates the path of the
mixture fraction as the system equilibrates. The instability level
is represented by the dashed horizontal line on the phase diagram.
The densities are in units of $g_0^2/a_{BB}^3$ and
$g_1^2/a_{BB}^3$ for bosons and fermions respectively, as defined
in Eq.~(\ref{eq:phases:nFB}). The inset (b) shows the sketch of
$E(\rho)$ at equilibrium for the corresponding phases. The inset
(c) gives the sketch of $E(\rho)$ near (below) the instability
line and (above) the phase separation curve. The arrow indicates
(schematically) modification of $E(\rho)$ during the
equilibration.}\label{fig1.eps}
\end{figure}

It was shown within Thomas-Fermi approximation \cite{Viverit} that
the equilibrium boson-fermion system with interactions introduced
above can exist in three configurations (phases): (i) mixture,
(ii) coexistence of pure fermion fraction with (spatially
separated) mixture, and (iii) coexistence of pure boson and pure
fermion fractions, see Fig.~\ref{fig1.eps}. In this section we
show that the structure of the entire phase diagram can be
obtained from the energy density (\ref{eq:intro:Eeff}). Indeed, we
notice that $E(\rho)$ has either one or two local minima. The
first one (if present) is at $\rho=0$. It clearly describes pure
fermion fraction with the density of fermion $\rho_F(0)$. The
position of the second minimum is found from $\partial_\rho
E(\rho_0)=0$. It describes either the mixture with fermion density
$\rho_F(\rho_0)>0$, or the pure boson fraction with
$\rho_F(\rho_0)=0$ due to the step function in
Eq.~(\ref{eq:phases:nF-gen}).

We first discuss the situation when the thickness of the
transition region (boundary) between the fractions, or the
coherence length, $l$, is microscopic (i.e. small with respect to
the characteristic size of the fractions). In this case it is
natural to introduce the overall (average) fermion $\rho_F =
\rho_F(0) V_F + \rho_F(\rho) V_B$ and boson $\rho_B=V_B\rho$
densities of a larger system. Here $V_F$ and $V_B=1-V_F$ are
relative volumes for each fraction.

When the densities $\rho_B$ and $\rho_F$ are small, the minimum at
$\rho=\rho_0$ is the only one present. At low temperatures the
equilibrium system will occupy this minimum creating uniform
mixture (phase i) with $\rho_F = \rho_F(\rho_0)$ and
$\rho_B=\rho_0$ ($v_F=0$). At higher densities the minimum at
$\rho=0$---the pure fermion fraction---forms. As a result, in
equilibrium, the two minima align, $E(0) = E(\rho_0)$, which
defines the volume of the spatially separated  fermion fraction.

For clarity of further discussion it is convenient to introduce
dimensionless densities \cite{Viverit} as
%////////////////////////////////////
\begin{equation}\label{eq:phases:nFB}
n_F \equiv \rho_Fa_{BB}^3/g_1^2%
\quad \mathrm{and} \quad %
n_B \equiv \rho_{(B)}a_{BB}^3/g_0^2.
\end{equation}
%////////////////////////////////////
Here and throughout the paper we adopt the convention of using
``n" for dimensionless densities corresponding to original
densities denoted by ``$\rho$". The conversion parameters of
Eq.~(\ref{eq:phases:nFB}) are
%////////////////////////////////////
\begin{equation}\label{eq:phases:g0}
g_0^2 = %
\frac{4a_{BB}^3\lambda_{BB}^2}{25\lambda_0^2\lambda_{BF}^5}=%
\frac{9\pi}{4} \frac{a_{BB}^5}{a_{BF}^5}\frac{m_F^2}{m_B^2}
\left(\frac{m_F}{m_B}+1\right)^{-5}
\end{equation}
%////////////////////////////////////
and
%////////////////////////////////////
\begin{equation}\label{eq:phases:g1}
g_1^2 = %
\frac{4a_{BB}^3\lambda_{BB}^3}{25\lambda_0^2\lambda_{BF}^6}=%
\frac{9\pi}{2} \frac{a_{BB}^6}{a_{BF}^6} \frac{m_F^3}{m_B^3}
\left(\frac{m_F}{m_B}+1\right)^{-6}
\end{equation}
%////////////////////////////////////
where $g_0$ is related to boson gas parameter (defined
\cite{LifshitzPitaevskii} via $g_B=\sqrt{\rho_B a_{BB}^3}$) as
$g_B^2 = g_0^2 n_B$.

The separation between phases i and ii on the phase diagram ($n_F$
vs $n_B$) occurs when $V_F$ sets to zero. The corresponding phase
separation curve can be easily obtained in a parametric form as
follows. Introducing $A\equiv\lambda_{BB}/5\lambda_0\lambda_{BF}^2
\mu_F^{1/2}$ and $y^2 \equiv 1-\rho_0\lambda_{BF}/\mu_F$ one can
write the boson and fermion densities in the mixture fraction as
%////////////////////////////////////
\begin{equation}\label{eq:phases:nFB-Ay}
n_F^0=\frac{y^3}{8A^3}%
\quad \mathrm{and} \quad %
n_B^0=\frac{1-y^2}{4A^2}.
\end{equation}
%////////////////////////////////////
The equilibrium mixture coexists with unoccupied ($V_F\to 0$) pure
fermion minima only at the i-ii phase transition curve. In terms
of $A$ and $y$, the equation $E(0) = E(\rho_0)$ takes the form
%////////////////////////////////////
\begin{equation}\label{eq:phases:yA}
A = \frac{2 + 4y + 6y^2 + 3y^3}{5(1 + y)^2}
\end{equation}
%////////////////////////////////////
which, together with Eq.~(\ref{eq:phases:nFB-Ay}) defines the
phase transition curve: $n_F^0$ vs $n_B^0$. From
Eq.~(\ref{eq:phases:yA}) we also see that $A$ varies within
$2/5\leq A \leq 3/4$ (since by definition $0\leq y \leq 1$ within
phase ii).

The complete separation on pure fermion and pure boson fractions
(phase iii) occurs for higher densities when $\rho_0 \geq
\mu_F/\lambda_{BF}$. This situation is not considered in the
present paper.

In the discussion above, it was assumed that in the phase ii the
fermion fraction is spatially separated from the mixture and the
volume of the surface layer is negligible. This is not always the
case as will be demonstrated in the next section. However when $l$
is large, one can still define effective partial volumes $V_F$ and
$V_B$, assuming that the densities discussed above are the peak
densities. While the relation between $V_F$ and $V_B$ is now
complicated, the limit $V_F\to 0$ with $V_B\to 1$ still exists.
Therefore, the i-ii separation curve found above is valid.

\section{Dynamics of Quantum Transition}\label{sec:equil}

We are interested in the kinetics of the phase transition between
phases i and ii, which is manifested by separation of pure fermion
fraction out of the mixture. When the uniform mixture is prepared
with the densities $n_F$ and $n_B$ above the i-ii separation
curve, e.g. point X, it occupies the second minima of $E(\rho)$
and is metastable, see Fig.~\ref{fig1.eps}c. As $n_F$ increases
towards the point Y (and further) the barrier that separates local
minima $\rho_0$ becomes smaller and eventually disappears from
$n_F \geq n_F^s$, with $n_F^s = 8/27$ (the absolute, spinodal,
instability line). From this point the mixture becomes unstable.

The decay of the metastable mixture results in spatial separation
of pure fermion fraction. At low temperatures this transition is
governed by quantum tunneling (in contrast to high temperatures,
when thermal activation becomes effective and transition is due to
thermal fluctuations rather then tunneling). To illustrate the
equilibration process, let us assume that the coherence length,
$l$, is small enough so that the spatial separation of the fermion
fraction is well defined. Towards the end of this section we will
show where such condition is met. During the equilibration, the
relative volume $V_F$ of the fermion fraction increases from 0 to
some equilibrium value $V_F^0$. The densities $\rho_0$ and
$\rho_F(\rho_0)$ of the remaining mixture will change, as well as
$E(\rho)$. The point $(\rho_0,\rho_F(\rho_0))$ will drift towards
i-ii separation curve where $E(0)=E(\rho_0)$, as shown by the
arrow in Figs.~\ref{fig1.eps}a and~\ref{fig1.eps}c. As it can be
easily verified, the drift line is parallel to ii-iii separation
line and its intersection with the $n_F$ axis gives the
equilibrium value for the density in pure fermion fraction,
$\rho_F(0)$. The partial volume of the later is $V_F^0 =
(n_B^0-n_B)/n_B^0$.

In what follows we will obtain the density profile for the fermion
droplet in various regions of the phase diagram (within phase ii)
and set up equations to describe the transition rate in the
system. The rates are calculated in the next sections.

The dynamics of the boson condensate due to tunneling part of the
equilibration is given by the transition amplitude $\bra{ii'}
e^{-iHt} \ket{i'} = \int \mathcal{D}\psi_B \mathcal{D}\psi_B^*
e^{i\int dt d\mathbf{r} L(\psi_B ,\psi_B^*)}$, where the rhs is
Feynman's sum over the histories \cite{Kleinert,Coleman}. The
state $\ket{i'}$ represent the (non-equilibrium) metastable
mixture residing in the second minimum of $E(\rho)$, at
$\rho\to\rho_0$. The state $\ket{ii'}$ correspond to the state
after the tunneling with $\rho\to\rho_1$, $E(\rho_1)=E(\rho_0)$.
Beyond this point the system hydrodynamically equilibrates to the
state $\ket{ii}$ where the two minima get aligned, see
Fig~\ref{fig1.eps}b, curve ii. The Lagrangian density is defined
as $L(\psi_B ,\psi_B^*) = \psi_B^* i\hbar\partial_t \psi_B -
H(\psi_B ,\psi_B^*)$, where the Hamiltonian density contains the
kinetic $\hbar^2|\nabla\psi _B|^2/2m_B$ and potential $E(\rho)$
contributions, where the latter is considered within the
Thomas-Fermi approximation, see Eq.~(\ref{eq:intro:Eeff}).

As it has been pointed out earlier, approximation for $E(\rho)$,
i.e. Eq.~(\ref{eq:intro:Eeff}), does not account for the gradient
terms which arise due to inhomogeneity of the fermion subsystem
and add to $\hbar^2|\nabla\psi _B|^2/2m_B$ term of the Lagrangian
density. In other words, the approximation implies that
renormalization of the boson kinetic energy arising due to the
non-locality of the fermionic response function is relatively
small. A straightforward perturbative estimate \cite{Kleinert} to
the second order in $\lambda_{BF}$ yields the gradient term
correction to the Thomas-Fermi of the order $\sim
(m_F^{3/2}\!\lambda_{BF}^2 \mu_F^{1/2} /\hbar^3
k_F^2)(\nabla\rho)^2$. Comparing this term with the bare boson
kinetic energy, we see that near the phase transition line it is
smaller by the factor of
$\sim(m_B\!/m_F)^{2/5}g_B^{2/5}n_B^{4/5}\!\!/n_F^{2/3}$.
Therefore, for sufficiently low $g_B$ and not very small fermion
densities the renormalization correction is negligible. This is
clearly the case in the region near the spinodal instability line.
In the vicinity of the phase separation line the renormalization
correction can become important (e.g. closer to the tricritical
point where $n_F\to 0$). In this case the correction is given by
the factor of $\sim(k_F l)^{-2}$, where the coherence length $l$
is of the order $a_{BB}/g_B$ or greater (as will be demonstrated
shortly). For $g_B\lesssim 0.1$ and not too small $n_F$ (e.g. for
$n_B\sim 0.4$), $(k_Fl)^2\sim 50$, and thus the Thomas-Fermi
approximation is well justified.

Considering the above arguments, the transition can be described
by the effective Lagrangian density of the form
%/////////////////////////////////////////////////////
\begin{equation}\label{eq:equil:L}
L  =  \hbar \rho \frac{d}{dt} \phi  + \frac{\hbar^2\rho}{2m_B}
\left(\nabla\phi\right)^2 + \frac{\hbar^2}{2m_B}
\frac{\left(\nabla\rho\right)^2}{4\rho} + E(\rho).
\end{equation}
%/////////////////////////////////////////////////////
Here the first term can be viewed as the Berry phase; the second
and the third terms arise due to the kinetic energy; the
dissipative terms due to particle-whole excitations in the fermion
subsystem will be included later. The contribution due to
non-condensate component of bosonic system is already accounted
for by the above treatment and therefore separate consideration is
not necessary for the purpose of the present calculation.

The decay rate from a metastable state can be obtained
\cite{Coleman} by calculating the classical action for the
transition amplitude in imaginary time formalism, $it\to t$.
Namely,
%/////////////////////////////////////////////////////
\begin{equation}\label{eq:equil:Gamma}
\Gamma/V = (\Gamma_0/V)\exp(-S/\hbar)
\end{equation}
%/////////////////////////////////////////////////////
where the action $S = \int
dtd\mathbf{r}L(\rho_\ind{cl},\phi_\ind{cl})$ is evaluated over the
classical (extremal) trajectory, $\rho_\ind{cl}(\phi_\ind{cl})$.
As will be shown below, for the parameters of interest, the rate
is dominated by the exponent and is less sensitive to the changes
in prefactor $\Gamma_0/V$. Therefore precise evaluation of
$\Gamma_0/V$ is not crucial. It can be estimated as
$\Gamma_0/V\sim \omega_0 l^3$, where $l$ is boson coherence length
as before, and $\omega_0$ is an ``attempt'' frequency. From the
uncertainty principle $\omega_0\sim \hbar/2m_B l^2$ and thus
$\Gamma_0/V\sim \hbar/m_Bl^5$.

The extremal action $S$ is calculated by setting the corresponding
functional derivatives to zero, i.e. $\delta L/\delta \rho =0$ and
$\delta L/\delta \phi =0$. The second derivative is trivial. It
leads to the continuity equation
%/////////////////////////////////////////////////////
\begin{equation}\label{eq:equil:continuityEQ}
\partial_t\rho+\nabla(\rho{\bf{u}})=0
\end{equation}
%/////////////////////////////////////////////////////
with $\mathbf{u} = \hbar\nabla\phi/m_B$ and can be used to
eliminate $\phi$. Equation.~(\ref{eq:equil:continuityEQ}) can be
easily solved for the velocity of the condensate assuming
spherical symmetry. The result is $\mathbf{u} =
({\mathbf{\hat{r}}}/{r^2\rho})\int_0^r dr{r^2
\partial _t \rho}$. After a straightforward algebra the action
becomes
%/////////////////////////////////////////////////////
\begin{eqnarray}\label{eq:equil:S}
S   = 4\pi\int dtdrr^2 \left[\frac{m_B}{2\rho} \left(\frac{1}{r^2}
\int_0^r drr^2\partial_t\rho\right)^2 \right.
\\ \nonumber
\left.+\frac{\hbar^2}{2m_B}\frac{\left(\nabla\rho\right)^2}{4\rho}
+ E(\rho)\right].
\end{eqnarray}
%/////////////////////////////////////////////////////

The pure fermion fraction to be formed during the equilibration
corresponds to the bubble in the boson system. Its shape is
defined by the last two terms of Eq.~(\ref{eq:equil:S}), i.e. the
equation $\delta S/\delta \rho =0$ in the static case,
%/////////////////////////////////////////////////////
\begin{equation}\label{eq:equil:difEqStatic}
\partial_r^2\sqrt\rho  + \frac{2}{r}\,\partial_r\sqrt\rho =
\frac{m_B}{\hbar^2}\,\partial_{\sqrt\rho} E(\rho)
%2\partial_r^2\sqrt\rho  + \frac{4}{r}\,\partial_r\sqrt\rho =
%\frac{2m_B}{\hbar^2}\,\partial_{\sqrt\rho} E(\rho)
\end{equation}
%/////////////////////////////////////////////////////
To solve it, let us first ignore the the second term of the
left-hand side, which is a reasonable approximation for large
enough bubbles. The solution of the remaining equation can be
easily estimated for all relevant area of the pase diagram, see
Appendix~\ref{app:shape}. Near the i-ii separation curve we obtain
%/////////////////////////////////////////////////////
\begin{equation}\label{eq:equil:rho_i-ii}
\rho(r) \sim \frac{\rho_0}{1+\exp[-4(r-R)/l_0]}
\end{equation}
%/////////////////////////////////////////////////////
where the coherence length, $l\to l_0$ is
%/////////////////////////////////////////////////////
\begin{equation}\label{eq:equil:L0}
\frac{l_0}{a_{BB}}\sim \frac{1}{g_0 n_B}
\end{equation}
%/////////////////////////////////////////////////////
Near the spinodal instability the first-derrivative term becomes
important. The density $\rho(r)$ varies between $\rho_1$ and
$\rho_0$. The coherence length $l\to l_s$ is found, see
Appendix~\ref{app:shape}, in the form
%/////////////////////////////////////////////////////
\begin{equation}\label{eq:equil:Ls}
\frac{l_s}{a_{BB}} = \frac{\sqrt{3}}{g_0\sqrt{\pi n_B}}
\left(1-\frac{n_F}{n_F^s}\right)^{-1/2} + {\cal
O}\left(1-\frac{n_F}{n_F^s}\right)
\end{equation}
%/////////////////////////////////////////////////////
The estimate of $l$ for all the densities of phase ii below
instability is given in Fig.~\ref{fig2.eps}a (see
Appendix~\ref{app:shape} for derivations). For not too small $g_0$
the characteristic length associated with the transition region is
microscopic, i.e. $l\sim a_{BB}$. This is the case for large part
of the interval $n_F^0\leq n_F<n_F^s$, see Fig.~\ref{fig2.eps}a.
Therefore formation of a distinct fermion fraction with thin
boundary (nucleation) is expected for this range of parameters. It
is straightforward to obtain the characteristic (critical) radius,
$R_c$ of such a nuclei. Integrating both sides of
Eq.~(\ref{eq:equil:difEqStatic}) with respect to $\sqrt{\rho}$ we
obtain $m_B\Delta E/\hbar^2 =
2\int_0^{\rho_0}d\sqrt{\rho}\partial_r\sqrt{\rho}/r$. When $l\ll
R$, we have $R_c  = \sigma/\Delta E$, where $\sigma \equiv
(2\hbar^2/m_B)\int_0^{\rho_0} d\sqrt{\rho}\partial_r\sqrt{\rho}$
is the surface tension.
\begin{figure}
\includegraphics[width=5.0cm]{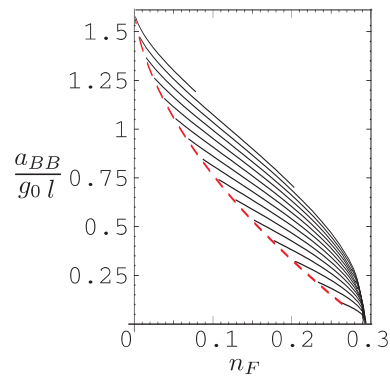}
\caption{(a) The reciprocal coherence length as a function of
dimensionless fermion density, $n_F$, for different values of
dimensionless boson density of the mixture: $n_B=0.1,0.2,...,1.5$
(from bottom to top). The dashed curve represents reciprocal
coherence length at the i-ii phase separation
curve.}\label{fig2.eps}
\end{figure}

\section{Effect of dissipation}\label{sec:dissipation}

We now consider the effect of dissipation which was ignored so far
in our description of the kinetics. Excitation of the
particle-hole pairs in the fermion subsystem interacting with the
condensate affects the transition, as it happens for other systems
with quantum tunneling
\cite{CaldeiraLeggett,DSVB,DMmosfet,DSmosfet,VBDS}. As will be
demonstrated further, the resulting dissipation is important and
can lead to significant modifications of the transition rate.

The dissipation terms arise naturally from more accurate treatment
of the boson-fermion interaction. Indeed, the Thomas-Fermi
approximation utilized in the derivation of the effective
potential $E(\rho)$, see Secs.~\ref{sec:model} and
\ref{sec:equil}, implies that fermions instantaneously (on a much
shorter time scale) adjust to the local variation of boson
density: it ignores the $\omega$-dependent terms in the
diagrammatic expansion of the partition function. The effect can
be analyzed by considering the second order correction produced by
the interaction $\lambda_{BF}\psi_F^\dag\psi_F\rho$.
%/////////////////////////////////////////////////////
\begin{equation}\label{eq:equil:DEC-vac-pol}
\frac{{\lambda _{BF}^2 }}{{2\hbar ^2 }}\int
{\frac{{d{\bf{q}}}}{{\left( {2\pi } \right)^3 }}\frac{{d\omega
}}{{2\pi }}\left( {\frac{{\hbar ^2 k_F^3 }}{{4\pi ^2 \mu _F }} +
\frac{{m_F^2 \left| \omega  \right|}}{{4\pi \hbar ^2 q}} +... }
\right)\left| {\rho (q,\omega )} \right|^2 }.
\end{equation}
%/////////////////////////////////////////////////////
The first $\omega$ independent term is already present in
Eq.~(\ref{eq:intro:Eeff}). We are after the frequency dependent
term $m_F^2|\omega|/4\pi\hbar^2q$ responsible for Landau damping.

In the limit $l\ll R$, i.e. near the phase separation curve and
far enough from the instability line, we can use
$\rho(r)\approx\rho_0\theta(r-R)$ as suggested by
Eq.~(\ref{eq:equil:rho_i-ii}). Substitution into
Eq.~(\ref{eq:equil:DEC-vac-pol}) gives the correction to $S$ in
the form
%/////////////////////////////////////////////////////
\begin{eqnarray}\label{eq:equil:DeltaS}
&&\Delta S \to \frac{\gamma_0}{64\pi} {\cal P}\int {dt dt'\over
(t-t')^2}\left\{R(t)^3R(t') \phantom{\frac{R}{R}} \right.
\\ \nonumber
&&\left. + R(t)R(t')^3 + \frac{1}{2}[R(t)^2 - R(t')^2]^2
\ln\left|\frac{R(t)-R(t')} {R(t)+R(t')}\right|\right\}.
\end{eqnarray}
%/////////////////////////////////////////////////////
Here $\gamma_0={4\lambda_{BF}^2 m_F^2 \rho_0^2}/{\pi \hbar^3}$.
The first two terms in the right-hand side arise due to the
restructuring of the fermionic density of states inside the
droplet in the course of its expansion, while the last term can be
viewed as coupling between droplet's surface and particle-hole
excitation in Fermi sea.

Near spinodal instability the tunneling barrier is small and the
correction is due to small variation of the density
$\rho=\rho_0+\delta\rho$ around the metastable minima, thus
%/////////////////////////////////////////////////////
\begin{equation}\label{eq:equil:DeltaS-spinodal}
\Delta S \to \frac{\lambda _{BF}^2}{2\hbar^3}\int
\frac{d{\bf{q}}}{(2\pi)^3} \frac{d\omega}{2\pi}
\frac{m_F^2|\omega|}{4\pi q} |\delta\rho(q,\omega)|^2.
\end{equation}
%/////////////////////////////////////////////////////

\section{Nucleation near the phase transition line}\label{sec:i-ii}

\begin{figure}
\includegraphics[width=5.0cm]{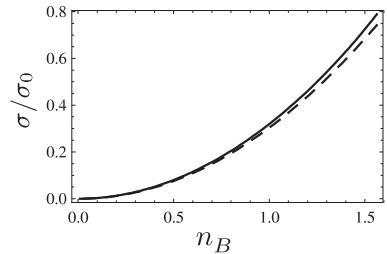}
\caption{Dimensionless surface tension coefficient along the phase
separation curve (numerical solution), $\sigma_0 =
\sqrt{\pi}\hbar^2g_0^3/2m_Ba^4_{BB}$. Approximation,
Eq.~(\ref{eq:i-ii:sigma-aprox}), is shown by the dashed
curve.}\label{fig3.eps}
\end{figure}

In this section we evaluate the i-ii transition rate near the
phase separation curve, see Fig.~\ref{fig1.eps}a. Evaluation of
the extremum action with the structure of Eq.~(\ref{eq:equil:S})
has been address in Ref.~\onlinecite{LifshitzKagan}. In the
vicinity of the transition curve the coherence length, $l\to l_0$,
is given by Eq.~(\ref{eq:equil:L0}). At the same time $\Delta E\to
0$ as we approach the i-ii transition curve (the two minima of
$E(\rho)$ align) and, therefore, $R_c\to \infty$. As the result
$l_0\ll R_c$ and the boson density entering
Eq.~(\ref{eq:equil:rho_i-ii}) can be approximated by a step
function $\rho(r,t) \approx \rho_0 \theta[r-R(t)]$ (the thin wall
approximation), where $\rho_0$ is bosonic density of the mixed
phase as before. Within such an approximation the action $S$ can
be formulated \cite{LifshitzKagan} in terms of $R(t)$. Evaluating
the integral in Eq.~(\ref{eq:equil:S}) over $r$ we obtain
%/////////////////////////////////////////////////////
\begin{equation}\label{eq:i-ii:S0}
S  = 4\pi \!\!\! \int dt\left[\frac{m_B\rho_0}{2}R^3
\left(\partial_tR\right)^2\!\! + \sigma R^2\! - \frac{\rho_0
\Delta\mu}{3} R^3 \right],
\end{equation}
%/////////////////////////////////////////////////////
Here $\sigma$ is the surface tension defined earlier. In the
present case (since $l\ll R$), the second term of
Eq.~(\ref{eq:equil:difEqStatic}) is negligible and a simple energy
conservation equation $\frac{\hbar^2}{2m_B}(\partial_r\sqrt\rho)^2
+E(\rho)=const$ corresponding to the static part of action $S$
holds. Therefore we can rewrite the surface tension as
%/////////////////////////////////////////////////////
\begin{equation}\label{eq:i-ii:sigma-def}
\sigma =\sqrt{\frac{\hbar^2}{2m_B}} \int\limits_{0}^{\rho_0}
d\sqrt{\rho}\sqrt{E(\rho)-E(\rho_0)}
\end{equation}
%/////////////////////////////////////////////////////
This integral is evaluated in Appendix~\ref{app:sigma}. The
numerical result is shown in Fig.~\ref{fig3.eps}. It can be
approximated by
\begin{equation}\label{eq:i-ii:sigma-aprox}
\sigma \approx 0.304\frac{\sqrt\pi\hbar^2}
{2m_Ba_{BB}^4}g_0^3n_B^2
\end{equation}
with the error of a few percents along the entire curve, see
Appendixes~\ref{app:shape} and \ref{app:sigma}.

It is expected that the surface tension coefficient will decrease
as one departs from the phase separation curve. However near the
transition line this change is not significant compared to the
change in the bulk energy. Therefore to estimate nucleation rates
it is sufficient to use Eq.~(\ref{eq:i-ii:sigma-aprox}). The bulk
energy is given by the integral of $E(\rho)$ over the volume of
the bubble. Therefore $\rho_0\Delta\mu$ is the energy difference
$\Delta\mu\rho_0\equiv E(\rho_0)-E(0)$ or the degree of
metastability, see Fig.~\ref{fig1.eps}c. When $n_F\to n_F^0$,
$\Delta\mu$ vanishes and therefore along the large part of the
phase separation curve $\Delta\mu\sim \Delta n_F$. However, this
in not the case near the end of the curve when $n_F^0\to 0$.
Generally,
%/////////////////////////////////////////////////////
\begin{equation}\label{eq:i-ii:delta-mu}
\frac{\Delta\mu a_{BB}^3}{\lambda_{BF}g_1^2} = K\Delta n_F + K'
\left[\frac{3}{2}(n_F^{2/3}n_{F,2}^{0\,1/3}-n_F^{0}) - \Delta
n_F\right]
\end{equation}
%/////////////////////////////////////////////////////
where
%/////////////////////////////////////////////////////
\begin{equation}\label{eq:i-ii:K}
K'=\frac{2}{3}\frac{(n_F^{0\,2/3}+n_B)^{3/2}}{n_Bn_F^{0\,1/3}}
\quad\quad K=K'-\frac{2}{3}\frac{n_F^{0\,2/3}}{n_B} - 1
\end{equation}
%/////////////////////////////////////////////////////
One can see that when $\Delta n_F\ll n_F^0$ the second term in
Eq.~(\ref{eq:i-ii:delta-mu}) vanished and we can use $\Delta\mu
\sim K\Delta n_F$ as pointed out earlier. In the opposite case
$K\sim K'\to 2n_B^{1/2}/3n_F^{0\,1/3}$ and $\Delta\mu \sim
n_B^{1/2}\Delta n_F^{2/3}$. In this last limit, however, one has
to account for renormalization of the boson kinetic term and,
thus, the surface tension coefficient $\sigma$, as discussed
earlier.

The phase transition and formation of the fermion droplet is due
to the interplay between the surface tension and bulk energy of
the droplet [the last two term in $S$, see
Eq.~(\ref{eq:i-ii:S0})], which creates a potential barrier. The
system (instanton) has to tunnel through this barrier creating the
critical droplet of radius $R_c=3\sigma/\rho_0\Delta\mu$. The rate
of such the nucleation process is found from $\delta S/\delta
R=0$, with the result
$\ln\Gamma/\Gamma_0=-5\pi^2\sqrt{m_B\rho_0\sigma}R_c
^{7/2}/2^{9/2}\hbar^2$. In terms of dimensionless densities of
Fig.~\ref{fig1.eps}a, for $\Delta n_F\ll n_F^0$ we obtain
%/////////////////////////////////////////////////////
\begin{equation}\label{eq:i-ii:GammaX}
\ln{\Gamma'/\Gamma_0}=-0.0056 \frac{n_B^{11/2}}{g_B
K^{7/2}}\left({n_F^s\over \Delta n_F}\right)^{7/2}.
\end{equation}
%/////////////////////////////////////////////////////
Here $g_B$ is the conventional boson gas parameter as define
earlier.

As expected, the tunneling exponent, i.e. the rhs of
Eq.~(\ref{eq:i-ii:GammaX}), is singular in the degree of
metastability $\Delta n_F$ and diverges as $\Delta n_F^{-7/2}$.
Equation (\ref{eq:i-ii:GammaX}) also indicates that the rate of
nucleation is exponentially small in the dilute limit, i.e., for
$g_B\ll 1$. Since the thin wall approximation (nucleation)
requires sufficiently high energy barrier, e.g.
Fig.~\ref{fig1.eps}(b,c), the rhs of Eq.~(\ref{eq:i-ii:GammaX})
can not be reduce significantly decreasing $n_B$, see
Eq.~(\ref{eq:equil:L0}) and Fig.~\ref{fig2.eps}a. However, due to
the smallness of the numerical coefficient one can hope that
quantum nucleation is observable in sufficiently strongly coupled
systems (which are presently realizable with the use of Feshbach
resonance). Indeed, for $g_B\sim 0.1$, $n_B\sim 0.4$, and $\Delta
n_F/n_F^s = 0.15$, the coefficient $K \sim 0.27$ and the tunneling
exponent, is $\sim -27$. For the same parameters and $a_{BB}\sim
20$ a.u. the estimate of the prefactor $\Gamma_0/V$ gives
$10^{11}$ s$^{-1}\mu$m$^{-3}$. As the result the nucleation rate
$\Gamma/V$ is of the order $1$ s$^{-1}\mu$m$^{-3}$. This is
readily observable.

Let us now include dissipation terms (\ref{eq:equil:DeltaS}).
Exact evaluation of extremal action $S+\Delta S$,
Eqs.~(\ref{eq:i-ii:S0}) and (\ref{eq:equil:DeltaS}), is not
possible and we use the variational technique. A natural anzats is
$R(t)=pR_c e^{-\alpha t^2}$, where coefficients $\alpha$ and $p$
are variational parameters. The expression for $\Delta S$,
Eq.~(\ref{eq:equil:DeltaS}), does not depend on the time scale of
$R(t)$. Rescaling of $R(t)$ yields $\Delta S\sim p^4 R_c^4$. The
proportionality coefficient depends only on the shape of the trial
function,\com{see Appendix~\ref{app:i-ii-ap},} and, thus, can be
estimated separately. The system $\partial_p(S+\Delta S)=0$,
$\partial_\alpha S=0$ leads to the solution
%/////////////////////////////////////////////////////
\begin{equation}\label{eq:i-ii:p}
p = \frac{{75\sqrt 6 \gamma ^2  + 42\sqrt {10} \pi  + 15\gamma
\sqrt {150\gamma ^2  + 7\sqrt {15} \pi } }}{{300\gamma ^2 +
32\sqrt {15} \pi }}
\end{equation}
%/////////////////////////////////////////////////////
where for $\Delta n_F\ll n_F^0$
%/////////////////////////////////////////////////////
\begin{equation}\label{eq:i-ii:gamma}
\gamma = 0.742 g_0^{1/5}n_B\left(\frac{m_F}{m_B}\right)^{4/5}
\sqrt{\frac{n_F^s}{K\Delta n_F}}
\end{equation}
%/////////////////////////////////////////////////////
Note that $p\to 7\sqrt 3/8\sqrt 2$ for $\gamma\to 0$ and $p\to
\sqrt{3/2}$ for $\gamma\to\infty$. The second limit gives the
leading asymptotic in the vicinity of the phase separation curve
%/////////////////////////////////////////////////////
\begin{equation}\label{eq:i-ii:GammaDEC}
\ln\Gamma''/\Gamma_0 \sim - 0.01
\frac{n_B^{32/5}}{g_B^{4/5}K^4}\left({m_F\over
m_B}\right)^{4/5}\left(n_F^s\over\Delta n_F\right)^4,
\end{equation}
%/////////////////////////////////////////////////////
The other limit gives expression (\ref{eq:i-ii:GammaX}) with
slight overestimate of the numerical prefactor due to variational
nature of the calculation. The overall solution is
%/////////////////////////////////////////////////////
\begin{equation}\label{eq:i-ii:GammaALL}
\frac{\ln\Gamma/\Gamma_0}{\ln\Gamma'/\Gamma_0}
=\!\left(\frac{8\sqrt 2}{7\sqrt 3 }\right)^{\!\!\!7/2}\!\!
\left[\!\sqrt{8}p^{7/2}\sqrt{1\!-\!\sqrt{\!\frac{2}{3}}p} +\!
\gamma\frac{5^{3/4}2^{1/4}}{\sqrt \pi}p^4\right]
\end{equation}
%/////////////////////////////////////////////////////
where we have rescaled the overall numerical coefficient to mach
expression (\ref{eq:i-ii:GammaX}) in the underdamped limit
$\gamma\to 0$. The exponent $\ln\Gamma/\Gamma_0$ as a function of
$\Delta n_F/n_F^s$ is shown in Fig.~\ref{fig4.eps}.

The crossover between power $4$ and $7/2$ asymptotic curves take
place at $\gamma\sim 1$, or
%/////////////////////////////////////////////////////
\begin{equation}\label{eq:i-ii:DECcrossover}
K\Delta n_F/n_F^s \sim g_B^{2/5} n_B^{9/5}
\left(\frac{m_F}{m_B}\right)^{8/5}
\end{equation}
%/////////////////////////////////////////////////////
Thus, we find that correction due to dissipation term, $\Delta S$,
strongly alters the tunneling exponent in the region $K\Delta
n_F/n_F^s\!\!\ll\!g_B^{2/5}n_B^{9/5}(m_F\!/m_B)^{8/5}$, while in
the opposite (non-dissipative) limit the tunneling exponent is
given by Eq.~(\ref{eq:i-ii:GammaX}). From
Eqs.~(\ref{eq:i-ii:GammaDEC}) and (\ref{eq:i-ii:DECcrossover}) we
conclude that the influence of dissipation is significant for
large $m_F/m_B$ mass ratio. From the crossover condition we also
see that the dissipative regime is realized near the phase
separation curve within a ``band" of width $\Delta n_F$ determined
by the relation (\ref{eq:i-ii:DECcrossover}). For mixtures with
small fermion-boson mass ratio this ``band" is relatively narrow
and for large enough fermion density the system stays in
non-dissipative regime, where the nucleation rate is given by
Eq.~(\ref{eq:i-ii:GammaX}).

Let us consider the previous example (with $g_B\sim 0.1$, $n_B\sim
0.4$, etc.) assuming $m_F=m_B$. For this set of parameters the
system is in the crossover region (at the edge of the dissipative
``band" on the phase diagram). Using Eq.~(\ref{eq:i-ii:GammaALL})
we obtain $\ln\Gamma/\Gamma_0 \sim -70$, which means that the
transition is not observable. Choosing a smaller mass ratio, e.g.
$m_F/m_B\sim 1/4$ (the right-hand side of
Eq.~(\ref{eq:i-ii:DECcrossover}) becomes smaller by an order of
magnitude), we shrink the dissipative region around the phase
transition curve. The system now is in ``quiet" (non-dissipative)
regime and the previous estimates for the rate obtained with
Eq.~(\ref{eq:i-ii:GammaX}) are valid. Therefore the dynamics of
quantum nucleation can be systematically observable only for cold
atom systems with sufficiently small mass ratio $m_F/m_B$.

\section{Tunneling near spinodal instability}\label{sec:spinodal}

In this section we evaluate the transition rate near the absolute
instability line, i.e. $n_F^s-n_F\ll n_F^s$. In this case the
barrier separating the metastable mixture becomes small, see
Fig.~\ref{fig1.eps}c, curve Y. To compute the tunneling part of
the transition based on action (\ref{eq:equil:S}) it is sufficient
to retain only a few terms in the expansion of $E(\rho)$ at
$\rho\to\rho_0$. The terms of the second and the third order in
$\delta\rho=\rho-\rho_0$ will form the barrier (we set
$E(\rho_0)=0$ for convenience). It is clear that in the kinetic
part of the action only the second order terms should be kept (the
third order terms containing gradients are small compared to the
third order potential terms). We obtain
%/////////////////////////////////////////////////////
\begin{eqnarray}\label{eq:spinodal:S}
S   = 4\pi\int dtdrr^2 \left[\frac{m_B}{2\rho_0}
\left(\frac{1}{r^2} \int_0^r dr^\prime
{r^\prime}^2\partial_t\delta\rho\right)^2 \right.
\\ \nonumber
\left.+\frac{\hbar^2}{8m_B}\frac{\left(\nabla\delta\rho\right)^2}{\rho_0}
+ a\delta\rho^2 +b\delta\rho^3\right],
\end{eqnarray}
%/////////////////////////////////////////////////////
where
%/////////////////////////////////////////////////////
\begin{equation}\nonumber
a = {2\pi\hbar^2 a_{BB}\over {3m_B}}\left( 1-{n_F\over
n_F^s}\right)
\end{equation}
%/////////////////////////////////////////////////////
and
%/////////////////////////////////////////////////////
\begin{equation}\nonumber
b={\hbar^2a_{BF}^5m_B\over 3a_{BB}m_F^2}\left(1+{m_F\over
m_B}\right)^5.
\end{equation}
%/////////////////////////////////////////////////////
The dependence of action (\ref{eq:spinodal:S}) on $n_F$ and $n_B$
can be obtained rescaling $r$, $t$, and $\delta\rho$, similarly to
Ref.~\onlinecite{LifshitzKagan}. By introducing dimensionless
variables \cite{Ls-note} $x=r\sqrt{8m_B\rho_0 a}/\hbar$, $\tau =
4\rho_0 a t/\hbar$, and $p=\delta\rho b/a$, the action $S$ can be
rewritten as $const\times s$, where $s[p(x,\tau)]$ is a
parameter-independent functional, the extremum of which is a
c-number. Its value can be estimated by variational anzats
$p=-p_0\exp{(-\alpha x^2 - \beta \tau^2)}$, where $\alpha$,
$\beta$ and $p_0$ are variational parameters. We find
$\alpha=\beta=4/3$, $p_0=9$, and $s\approx 8.95$. Upon a
straightforward calculation one obtains
%/////////////////////////////////////////////////////
\begin{equation}\label{eq:spinodal:Gamma}
\ln{\Gamma_Y/\Gamma_0}=-{0.324\over g_B n_B^2}\left(1-{n_F\over
n_F^s}\right)^{1/2},
\end{equation}
%/////////////////////////////////////////////////////
Again we see that tunneling exponent is controlled by the inverse
boson gas parameter $g_B$.

The exponent vanishes when fermion density $n_F$ reaches the
instability line $n_F^s$, where effective energy barrier
disappears, see Fig.~\ref{fig1.eps}c. In this case the transition
rate is defined by the prefactor $\Gamma_0/V$ rather then
exponent, which is especially troublesome for a dense system. This
limit, however, is not of interest here. Moreover, for not very
small $n_B$ the observable transition occurs much closer to the
phase separation curve (far from the instability point) and is
described by Eq.~(\ref{eq:i-ii:GammaALL}) of Sec.~\ref{sec:i-ii}.

When $g_B\lesssim 0.01$ and $n_B\sim 0.1-1$, the phase transition
is not observable near the transition curve, see
Eqs.~(\ref{eq:i-ii:GammaX}) and (\ref{eq:i-ii:GammaDEC}),---one
has to chose the values of $n_F$ closer to the spinodal
instability. In this case the exponent (\ref{eq:spinodal:Gamma})
is significant. It rises rapidly as the function of $n_F^s-n_F$
defining the transition rate near the instability region. In this
respect Eq.~(\ref{eq:spinodal:Gamma}) is similar to the results on
Macroscopic Quantum Tunneling (MQT) in systems of trapped bosons
with attractive interactions~\cite{UedaLeggett}. When the
coherence length $l_s$, see Eq.~(\ref{eq:equil:Ls}) is still much
smaller then the size of the condensate, the height of the
potential barrier and thus the tunneling exponent for the critical
droplet are controlled not by the total number of particles in the
trap, but the local densities. Therefore, for sufficiently large
numbers of particles in the trap the tunneling exponent can be
fine-tuned with a desired accuracy, which can make observation of
the MQT rate possible in a well controlled and predictable regime.

\begin{figure}
\includegraphics[width=8.1cm]{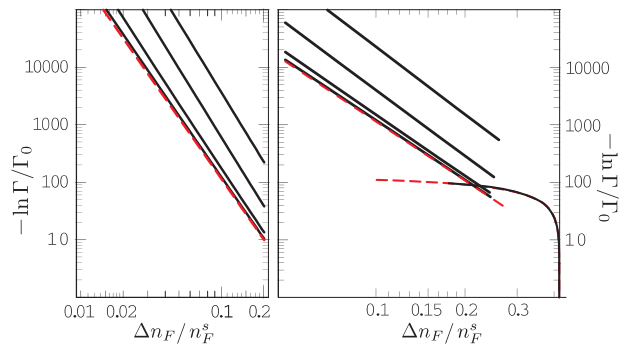}
\caption{Transition rate exponent, $\ln\Gamma/\Gamma_0$, as a
function of fermion density, $n_F=n_F^0+\Delta n_F$, for
$n_B=0.4$, $g_B=0.1$ (left) and $g_B=0.01$ (right). Asymptotics
near the i-ii phase separation curve are plotted for
$m_F/m_B=0.01,0.1,1,10$ (solid curves, bottom to top). The dashed
line represents the rate without dissipation, see
Eq.~(\ref{eq:i-ii:GammaX}). The asymptotic near the instability
region, Eq.~(\ref{eq:spinodal:Gamma}), is given in the right plot
(for the same variation of $m_F/m_B$).}\label{fig4.eps}
\end{figure}

The contribution of dissipation, e.g., action $\Delta S$ in
Eq.~(\ref{eq:equil:DeltaS-spinodal}), can be estimated by using
the same variational anzats as above. Rewriting the integral in
Eq.~(\ref{eq:equil:DeltaS-spinodal}) in terms of $x$, $\tau$ and
$p$ one finds that $\Delta S$ does not depend on $a$ and thus
$n_F-n_F^s$. Therefore for large enough $n_F-n_F^s$ and small
$g_B$ one can treat the dissipation perturbatively substituting
$p(x,\tau)$ from the above calculations. As the result the
tunneling exponent acquires an additional term $\sim -(g_B
n_B^2m_F/m_B)^{4/5}$. This term is again controlled by the mass
ratio $m_F/m_B$. For not too high fermion/boson mass ratio, it is
of the order $g_B$ and thus dissipation does not significantly
alter the dynamics of the phase transition (MQT) in this region,
see Fig.~\ref{fig4.eps}.

\section{Thermal nucleation}\label{sec:thermal}

It is instructive to analyze the influence of the classical
(thermal) activation mechanism. In order leave the metastable
state finite temperature system can (in addition to direct
tunneling leakage from its false ``ground" state) use its entire
spectrum tunneling from high energy states or going over the
barrier. The rate of such transition is proportional to the sum
\cite{LifshitzKagan,Gorokhov} of $\Gamma(E_n)\exp[-(E_n-E_0)/kT]$,
where the prefactor accounts for possible tunneling from $n$-th
exited state of the metastable mixture. This regime becomes
effective when the largest energy gap---between the first exited
and the ground state energies of the mixture---is of the order
$k_BT$. This energy difference enters the expression for quantum
nucleation rate and can be easily estimated in a similar manner
%/////////////////////////////////////////////////////
\begin{equation}\label{eq:expansion:E01}
E_1-E_0\sim \frac{p^2}{2m_B} \sim \frac{1}{2m_B}
\left(\frac{\hbar}{l}\right)^2.
\end{equation}
%/////////////////////////////////////////////////////
Here the characteristic length scale is the coherence length of
boson-fermion mixture.

Near the phase transition curve $l$ is given by
Eq.~(\ref{eq:equil:L0}), i.e. $l\sim a_{BB}/g_0 n_B$. This gives
the relation
%/////////////////////////////////////////////////////
\begin{equation}\label{eq:expansion:i-ii-T0}
g_B ^2 n_B\sim \frac{m_Ba_{BB}^2k_BT}{\hbar^2}
\end{equation}
%/////////////////////////////////////////////////////
Critical temperature of BEC transition can be estimate by that of
the uniform non-interacting three-dimensional boson gas
$k_BT_c\sim \hbar^2 n_B^{2/3}/m_B$, and we finally obtain
condition for thermal assisted transition in the form
%/////////////////////////////////////////////////////
\begin{equation}\label{eq:expansion:i-ii-Tc}
g_B^{2/3} n_B \sim \frac{T}{T_c}
\end{equation}
%/////////////////////////////////////////////////////
Our boson system is in deep BEC state, i.e. $T\ll T_c$, therefore
thermal fluctuations are not significant for observable quantum
nucleation that occurs in sufficiently dense systems ($n_B\sim
1$).

In dilute systems, where direct tunneling from low laying states
is extremely slow, one has two compute the entire sum
%/////////////////////////////////////////////////////
\begin{equation}\label{eq:expansion:Gamma-gen}
\Gamma = \sum_n\Gamma_0(E_n)e^{-S_n/\hbar-(E_n-E_0)/kT}
\end{equation}
%/////////////////////////////////////////////////////
up to the continuum states where $S_n\to 0$. The summation can be
carried out semiclassically minimizing the new energy functional,
i.e. exponent of Eq.~(\ref{eq:expansion:Gamma-gen}), see e.g.
Ref.~\onlinecite{LifshitzKagan}.

Closer to the spinodal instability the barrier becomes smaller and
the tunneling is effective even for relatively dilute systems. For
the crossover temperature we obtain
%/////////////////////////////////////////////////////
\begin{equation}\label{eq:expansion:spinodal-Tc}
g_B^{2/3}(1-n_F/n_F^s)\sim T/T_c.
\end{equation}
%/////////////////////////////////////////////////////
In this case, as expected, the zero-temperature MQT mechanism can
be observed in certain distance from the instability line, i.e.
when the left-hand side of Eq.~(\ref{eq:expansion:spinodal-Tc}) is
greater.

\section{Expansion of supercritical droplet}\label{sec:expansion}

As soon as a critical droplet is created as the result of
tunneling or temperature fluctuations, it should grow
equilibrating the system as pointed out in Sec.~\ref{sec:equil}.
Below we discuss only the case of finite coherent length---the
situation when a distinct stable droplet is created (thin-wall
approximation). We will estimate the expansion rate for $l\ll
R_c\ll R$ in the limits of large and small dissipation.

After a given nuclei passes (tunnel trough) the energy barrier,
its growth can be described by classical equation of motion for
radius $R$, $\delta S/\delta R + f_d =0$. Here $f_d$ is the
dissipative force (external time-dependent force is negligible at
low temperatures), and the action $S$ should be formulated in real
(original) time, i.e. in action (\ref{eq:i-ii:S0}) one replaces
$it\to t$. In the underdamped limit, $f_d \to 0$ (energy
conserved), the expansion is described by
%/////////////////////////////////////////////////////
\begin{equation}\label{eq:expansion:Econsrv}
%m_B\rho_0 \partial_t[R^3\partial_tR] = -2\sigma R +
%\rho_0\Delta\mu R^2.
\frac{{m_B \rho _0 R^3 }}{2}\left( {\partial _t R} \right)^2 =  -
\sigma R^2  + \Delta \mu \frac{{\rho _0 R^3 }}{3}.
\end{equation}
%/////////////////////////////////////////////////////
For large droplets ($R\gg R_c$) the surface tension term is
negligible and the expansion is linear in time $R \simeq \alpha_0
t$, with
%/////////////////////////////////////////////////////
\begin{equation}\label{eq:expansion:alpha_0}
\alpha_0 = \sqrt{2\Delta\mu/3m_B}.
\end{equation}
%/////////////////////////////////////////////////////

Fermion particle-hole excitations affect the the expansion
dynamics, renormalizing the expansion rate $\alpha_0$. To estimate
this effect we need to account for the dissipative part of the
action. This can be done by analytical continuation of Matsubara
action, Eq.~(\ref{eq:equil:DEC-vac-pol}), to Keldysh contour.
Following Ref.~\onlinecite{Kamenev} we obtain
%/////////////////////////////////////////////////////
\begin{equation}\label{eq:expansion:DeltaS}
\Delta S = \frac{m_F^2\lambda_{BF}^2}{2\pi\hbar^3} \!\!\int\!\!
dt\frac{d{\bf{q}}}{(2\pi)^3}\frac{1}{q}
\dot\rho_{cl}(q,t)\rho_q(q,t) + \O(\rho_q^2(q,t)),
\end{equation}
%/////////////////////////////////////////////////////
where $\rho_{cl/q}=(\rho_+\pm\rho_-)/\sqrt{2}$ and $\rho_\pm$
resides on forward (backward) branch. In the nucleation limit
(thin wall approximation) the time dependance of
$\rho(\mathbf{r},t)$ enters only via radius of the droplet,
$R(t)$, therefore $\rho_\pm=\rho\{R_\pm\}$. In the classical limit
we keep only terms linear in $R_q$. The dissipative force becomes
%/////////////////////////////////////////////////////
\begin{equation}\label{eq:expansion:ff}
f = \frac{m_F^2\lambda_{BF}^2}{2\pi\hbar^3} \!\!\int\!\!
dt\frac{d{\bf{q}}}{(2\pi)^3}\frac{1}{q}
\dot\rho_{cl}(q,t)\frac{\delta\rho_q(q,t)}{\delta R}.
\end{equation}
%/////////////////////////////////////////////////////
Substitution of $\rho(\mathbf{r},t)=\rho_0\theta(r-R(t))$ will
lead to the logarithmic divergence in the integration over the
momentum. The divergence is due to the high $q$. The cutoff comes
from the thickness of the surface area of the droplet. Using a
more accurate expression for $\rho(\mathbf{r},t)$, i.e.
Eq.~(\ref{eq:equil:rho_i-ii}), we obtain
%/////////////////////////////////////////////////////
\begin{equation}\label{eq:expansion:f}
f = - \gamma _e\gamma_0 \dot R(t)R(t)^2\ln\frac{4R(t)}{\pi l},
\end{equation}
%/////////////////////////////////////////////////////
where $\gamma_e=0.577...$ is the Euler constant, and $l$ is the
coherence length. As the result, expansion of a large
supercritical droplet is governed by
%/////////////////////////////////////////////////////
\begin{eqnarray}\label{eq:expansion:diss-eq}
m_B \rho _0 R\partial_t^2 R &+& \frac{3m_B \rho _0}{2}
(\partial_tR)^2
\\\nonumber
&+&\gamma _e\gamma_0(\partial_tR)\ln\frac{4R}{\pi l} - \rho _0
\Delta \mu=0.
\end{eqnarray}
%/////////////////////////////////////////////////////
This equation can be analyzed numerically for dimensionless
variables. Define $t=t_0\tau$, $R(t)=R_c\chi(\tau)$, then
Eq.~(\ref{eq:expansion:diss-eq}) becomes
%/////////////////////////////////////////////////////
\begin{equation}\label{eq:expansion:diss-eq-dless}
- \chi \partial_\tau^2 \chi  - \frac{3}{2}(\partial_\tau\chi) ^2 -
f_0(\partial_\tau\chi) \ln (\chi/{\xi}) + 1 = 0.
\end{equation}
%/////////////////////////////////////////////////////
Here $f_0 = \gamma _e\gamma_0 t_0/m_B\rho_0R_c$, $\xi = \pi
l/4R_c$, and $t_0^2 = R_c^2m_B/\Delta\mu$. Numerical integration
shows that for $f_0\gtrsim 1$ and $\tau\gtrsim 100 f_0$ the first
and the second terms remain smaller by several orders of
magnitude. Therefore we can ignore this terms in such the limit.
Moreover, we notice that logarithmic pre-factor of the third term
does not change significantly for large supercritical droplets.
Therefore with sufficient accuracy the growth is still linear in
time $R(t)\simeq \alpha t\ln^{-1}{(4\alpha t/\pi l)}$ with
%/////////////////////////////////////////////////////
\begin{equation}\label{eq:expansion:rate}
\alpha=\frac{\rho_0\Delta\mu}{\gamma _e\gamma_0},
\end{equation}
%/////////////////////////////////////////////////////

One can notice that $f_0\sim 1/\sqrt{\Delta\mu}$ and therefore
diverges at the phase separation curve. As the result, the above
solution is valid near the phase separation line. When the system
is placed farther away from the phase separation curve,
$\Delta\mu$ increases and $f_0$ becomes smaller. In this case we
have to retain the first two derivative terms in
Eq.~(\ref{eq:expansion:diss-eq-dless}). It changes the asymptotic
behavior. In the limit $f_0\to 0$ we arrive to
Eq.~(\ref{eq:expansion:Econsrv}) with the solution
$R(t)\backsimeq\alpha_0t$, where $\alpha_0$ is given by
Eq.~(\ref{eq:expansion:alpha_0}).

Similar to nucleation, the expansion of a supercritical droplet is
controlled by particle-whole excitations if the system is prepared
within a certain ``dissipative" region around the phase separation
line, e.g., Sec. \ref{sec:i-ii}. For larger degree of
supersaturation, dissipation is less effective. The ``dissipative"
region can be approximately estimated from condition
$f_0(\Delta\mu)\gtrsim 1$. We should note that measurement of the
growth rate does not necessarily require extremely dense systems
in which nucleation is observable. Supercritical droplet can be
created externally, e.g. by some laser fields or trapped atoms of
different sort. For such initiated supercritical droplet the
growth rate could be readily observable.

\section{Discussion}\label{sec:discus}

In the previous sections we have considered dynamics of the nuclei
of a new phase (i.e., pure fermion phase) in the fermion-boson
mixture in the metastable state. We have derived asymptotic
expressions for the nucleation rates for the system in two
regimes: in the regime of weak metastability, e.g., near the phase
transition line (Sec.~\ref{sec:i-ii}) and near the spinodal
(absolute) instability.

Analysis of Sec.~\ref{sec:i-ii}
leads to the conclusion that in the vicinity of the phase transition line
the nucleation dynamics is practically not observable for sufficiently dilute systems.
e. g., with $g_B\ll 1$.   It may appear that the dependence of Eqs.~(\ref{eq:i-ii:GammaX}),
(\ref{eq:i-ii:GammaDEC}), and (\ref{eq:i-ii:GammaALL}) on $n_B$
contradicts to this statement. Indeed the transition rate
decreases with $n_B$ if $\Delta n_F$ is kept constant. Such limit,
however, is incorrect, since the nucleation condition $l_0\lesssim
R_c$ in not satisfied, $l_0\sim 1/n_B$ (or $1/\sqrt{n_B}$ if $g_B$
is constant). Evaluating $R_c$ in terms of $\Delta n_F$ one
obtains the condition on $\Delta n_F$ for which the asymptotes
(\ref{eq:i-ii:GammaX}), (\ref{eq:i-ii:GammaDEC}), and
(\ref{eq:i-ii:GammaALL}) are valid. We have $R_c/a_{BB} =
0.219(n_B^{3/2}\!\!/g_BK)(n_F^s/\Delta n_F)$. In the limit $n_B\to
0$ it gives $\Delta n_F/n_F^s\lesssim n_B$ (where the densities
are dimensionless, i.e. in units of Fig.~\ref{fig1.eps}a, as
before). Therefore in $n_B\to 0$ limit the power $7/2$ and $4$
asymptotes are not measurable.

Measurable transition rates for dilute systems can be found closer
to the instability region. In this case one enters a crossover
between nucleation and MQT regimes. As an example, let us consider
the mixture with the boson gas parameter $g_B\sim 0.01$. For
$a_{BB}\sim 10$nm and about $10^{8}$ confined particles. This
corresponds to the densities of the order $10^{14}$cm$^{-3}$ with
size of the trapped condensate $\sim 100\mu$m. For this parameters
the position of the metastable mixture on the phase diagram,
Fig.~\ref{fig1.eps}a, is primarily defined by the ratio
$a_{BB}/a_{BF}$, see Eqs.~(\ref{eq:phases:g0}) and
(\ref{eq:phases:g1}), and can be controlled by changing the
scattering lengths. The transition rate is $\Gamma/V\sim
x^5\exp(-x/g_Bn_B^2)$ where $x^2=1-n_F/n_F^s$, see
Eqs.~(\ref{eq:spinodal:Gamma}), (\ref{eq:equil:Ls}). The change of
the prefactor is not significant due to fast decay of the exponent
form $x \gtrsim 0.1n_B^2$ (for smaller $x$ accurate evaluation of
$\Gamma_0/V$ is necessary). Therefore we conclude that for
$(n_F^s-n_F)/n_F^s \gtrsim 0.01 n_B^4$ and $n_B\lesssim 1$ the
dependence of the exponent (\ref{eq:spinodal:Gamma}) on the
densities is measurable. In this case the coherence length $l_s
\gtrsim 1\mu$m. It rises for small $n_F^s-n_F$, however at $l_s\ll
100\mu$m (or $1-n_F/n_F^s\gg 10^{-4}$) the transition is still
controlled by local densities.

We have also found, e.g. Sections \ref{sec:dissipation} and
\ref{sec:i-ii}, that dissipation can rather strongly alter the
dependence of the nucleation rate in the vicinity of the phase
transition line. While the results obtained in these sections are
likely to be correct only quantitatively, they show that in the
vicinity of the transition line excitation of particle-hole pairs
in the Fermi sea plays crucial role. As a result the Thomas-Fermi
approximation is inapplicable within the ``band'' controlled by
the fermion-boson interaction strength as well as fermion-boson
mass ratio; see Sec.~\ref{sec:i-ii}.

Finally we considered the dynamics of supercritical droplets, i.e,
the ones which have tunneled to the supercritical size. We have
found that in the non-dissipative regime the radius of the droplet
grows linearly with time with the expansion rate coefficient
proportional to $\sqrt{\Delta n_F}$. In dissipative region, which
occurs near the phase transition line, the situation is quite
similar: up to logarithmic corrections the radii of the droplets
still grow linearly with time, with rate now being dependent on
friction coefficient, see Sec.~\ref{sec:expansion}.

{\acknowledgements We thank Eddy Timmermans for valuable
discussions and comments. DS acknowledges stimulating discussions
with Vladimir Privman. The work is supported by the US DOE. }

%////////////////////////////////////////////////////////////
%////////////////////////////////////////////////////////////
%////////////////////////////////////////////////////////////
\appendix
\renewcommand{\theequation}{\thesection .\arabic{equation}}

\section{Density Profile}\label{app:shape}

To analyze the shape of the condensate density function $\rho(r)$
of the bubble, i.e. the solution of
Eq.~(\ref{eq:equil:difEqStatic}), it is convenient to introduce
dimensionless quantities as $x=r/{\frak r}$, $\xi=t/{\frak t}$,
$\varepsilon =E/{\frak E}$, $s=S/{\frak S}$,
$u=\rho\lambda_{BF}/\mu_F$, and write the action
(\ref{eq:equil:S}) in the form
%/////////////////////////////////////////////////////
\begin{equation}\label{eq:shape:s}
s = \int d\xi x^2 dx \left[ \frac{1}{u}\left(\frac{1}{x^2}
\int\limits_0^x x^2\partial_\xi udx \right)^2 +
\frac{\left(\partial_x u \right)^2}{u} + \varepsilon(u) \right]
\end{equation}
%/////////////////////////////////////////////////////
with
%/////////////////////////////////////////////////////
\begin{equation}\label{eq:shape:E}
\varepsilon(u) = -\left[5Au_0 + \frac{5}{2}(1-u_0)^{3/2} \right]u
+ \frac{5}{2}Au^2 - (1-u)^{5/2}
\end{equation}
%/////////////////////////////////////////////////////
where $u_0={\lambda_{BF}\rho_0}/{\mu _F} = 1-y^2$, ${\frak t} =
{\hbar}/{4\lambda_0\lambda_{BF}\mu_F^{3/2}}$, ${\frak r}^2 =
{\hbar {\frak t}}/{2m_B}$, and ${\frak E} = \lambda_0\mu_F^{5/2}$.
Note that the parameters $A$ and $y$ (or $u_0$), introduced in
Sec.~\ref{sec:phases}, are generally independent and define the
position on the phase diagram via the same relation as in
Eq.~(\ref{eq:phases:nFB-Ay}). The relation (\ref{eq:phases:yA})
between them defines the special case---the equilibrium state
(residing on the phase separation curve).

Here we are after the time-independent part of the action $s$. In
the current notation equation (\ref{eq:equil:difEqStatic}) reads
%/////////////////////////////////////////////////////
\begin{equation}\label{eq:shape:difEqStatic}
\partial_x^2\sqrt u + \frac{2}{x}\partial_x\sqrt u =
\frac{1}{4}\frac{\partial\varepsilon(u)}{\partial{\sqrt u }}
\end{equation}
%/////////////////////////////////////////////////////
Let us first discuss the case when the second term of the
left-hand side is negligible (which is clearly the case near the
phase transition curve). Equation~(\ref{eq:shape:difEqStatic})
can, then, be simplified to
%/////////////////////////////////////////////////////
\begin{equation}\label{eq:shape:DEq}
\partial_x u = \sqrt u \sqrt{\varepsilon(u)-\varepsilon(u_0)},
\end{equation}
%/////////////////////////////////////////////////////
noticing that $(\partial_x\sqrt u)\partial_{\sqrt u}=\partial_x$.
This is also the consequence of energy conservation at large $x$
corresponding to the static part of action $s$. The right-hand
side of (\ref{eq:shape:DEq}) is too complicated for the equation
to be solved analytically. However, for the purpose of
integration, $\varepsilon(u)-\varepsilon(u_0)$ can be approximated
in the range $u_1\leq u\leq u_0$, where
$\varepsilon(u_1)=\varepsilon(u_0)=0$, as
%/////////////////////////////////////////////////////
\begin{equation}\label{eq:shape:approxE}
\sqrt{\varepsilon(u)-\varepsilon(u_0)} \approx 4{\frak W}(u - u_1
)^{1/2} (u_0  - u)
\end{equation}
%/////////////////////////////////////////////////////
with
%/////////////////////////////////////////////////////
\begin{equation}\label{eq:shape:W}
{\frak W}^2 =
\frac{\varepsilon(u_m)-\varepsilon(u_0)}{4^2(u_m-u_1)(u_0-u_m)^2}
\end{equation}
%/////////////////////////////////////////////////////
Here $u_1<u_m<u_0$ and $u_m\sim (u_1+u_0)/2$. This approximation
is exact at $u=u_m$, i.e. close to the maximum of the right-hand
side of Eq.~(\ref{eq:shape:DEq}) and therefore should lead to
correct estimate of the characteristic length scale of $u(x)$. The
maximal absolute deviation $\eta(u)\!\equiv\! 1\!-\!4^2{\frak
M}^2(u\!-\!u_1)(u_0\!-\!u)^2\!/[\varepsilon(u)\!-\!\varepsilon(u_0)]$
occurs at $u\to u_0$ and $u\to u_1$ at the end of the i-ii
separation curve ($u_0\to 1$), as can be easily verified
numerically. At that point we obtain $\eta(u_0)\approx 0.421$. The
error vanishes at the limit of small barrier, i.e. near the
instability line on the phase diagram. With this approximation
Eq.~(\ref{eq:shape:DEq}) can be easily solved with the result
%/////////////////////////////////////////////////////
\begin{eqnarray}\label{eq:shape:solution}
&&\exp \left[ 4{\frak W}\sqrt{u_0 }\sqrt{u_0-u_1}(x-const)\right]
=
\\ \nonumber
&&= \frac{{\left( {\sqrt u + \sqrt {u_0 } } \right)\left( {\sqrt
{uu_0 }  + \sqrt {u - u_1 } \sqrt {u_0  - u_1 }  - u_1 }
\right)}}{{\left( {\sqrt u - \sqrt {u_0 } } \right)\left( {\sqrt
{uu_0 }  - \sqrt {u - u_1 } \sqrt {u_0  - u_1 }  - u_1 } \right)}}
\end{eqnarray}
%/////////////////////////////////////////////////////
Here the characteristic size of the bubble, $R$, enters into the
constant of integration. The characteristic length-scale on the
surface of the bubble (the coherence length) is
%/////////////////////////////////////////////////////
\begin{equation}\label{eq:shape:l}
l = \frac{{\frak r}}{{\frak W}u_0\sqrt{1-u_1/u_0}}
\end{equation}
%/////////////////////////////////////////////////////
where ${\frak r}/a_{BB} = \sqrt{5/4}A^{3/2}/\sqrt{2\pi}g_0$.

In the vicinity of \textit{the i-ii separation curve} $u_1\to 0$,
$A$ is a function of $u_0$ according to Eq.(\ref{eq:phases:yA}),
and we use $u_m=u_0/2$. Substitution to Eq.~(\ref{eq:shape:W})
yields
%/////////////////////////////////////////////////////
\begin{equation}\label{eq:shape:i-ii_W}
{\frak W}^2 = \frac{{1 - \left( {1 - u_0 /2} \right)^{5/2}
}}{{2u_0 ^3 }} - \frac{{15}}{{16}}\frac{A}{{u_0 }} -
\frac{5}{8}\frac{(1-u_0)^{3/2}}{u_0^2}
\end{equation}
%/////////////////////////////////////////////////////
As the function of $u_0$, ${\frak W}\sqrt{A}$ varies from
$~0.1195$ to $0.1212$ with the mean value of $0.1203$ (the
root-mean-square deviation is $0.0004$). Therefore near the i-ii
separation curve with the same precision the coherence length,
$l_0$, becomes
%/////////////////////////////////////////////////////
\begin{equation}\label{eq:shape:i-ii_l}
\frac{l_0}{a_{BB}} = \frac{1}{g_0 n_B}
\end{equation}
%/////////////////////////////////////////////////////

In the vicinity of \textit{spinodal instability} $u_0-u_1 \approx
2^5A^2(1-n_F/n_F^s)/9$ as can be verified by expanding
$\varepsilon$ in powers of $u_0-u_1$, and we use $u_m =
(u_0+2u_1)/3$. Substitution into Eq.~(\ref{eq:shape:W}) yields
%/////////////////////////////////////////////////////
\begin{equation}\label{eq:shape:spinodal_W}
{\frak W}^2 = \frac{5u_0}{256\sqrt{1-u_0}} + {\cal O}(u_0-u_1)
\end{equation}
%/////////////////////////////////////////////////////
Therefore the coherence length, $l_s$, clearly diverges as
$1/\sqrt{1-n_F/n_F^s}$. Collecting all dimensional factors we
obtain
\begin{equation}\label{eq:shape:spinodal_l}
\frac{l_s}{a_{BB}} = \frac{\sqrt{3}}{g_0\sqrt{\pi n_B}}
\left(1-\frac{n_F}{n_F^s}\right)^{-1/2} + {\cal
O}\left(1-\frac{n_F}{n_F^s}\right)
\end{equation}
%/////////////////////////////////////////////////////

In the above analysis we have ignored the first derivative term in
Eq.~(\ref{eq:shape:difEqStatic}). It is strictly speaking
negligible only near the phase transition curve for large
thin-boundary bubbles. In the instability region solution
(\ref{eq:shape:solution}) will change. Nevertheless, the analysis
of Eq.~(\ref{eq:shape:difEqStatic}) shows taht the characteristic
length scale still has to be inverse-proportional to $\frak{M}u_0$
and incorporate the divergence of $1/\sqrt{1-u_1/u_0}$. Therefore
relation~(\ref{eq:shape:l}) still holds. This can be demonstrated
by expanding the right-hand side of
Eq.~(\ref{eq:shape:difEqStatic}) in terms of $1-u_1/u_0$. In the
intermediate region, i.e. $0<u_0-u_1<u_0$, the order-of-magnitude
agrement is expected since no other features are present, as can
also be verified numerically.

\begin{figure}
\includegraphics[width=8.1cm]{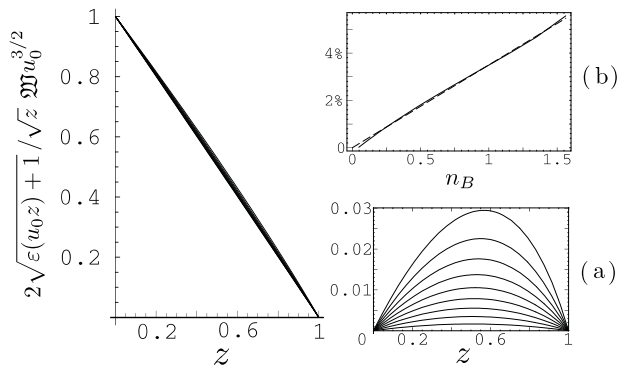}
\caption{Normalized energy deviation, see
Eq.~(\ref{eq:app:sigma:int}), for $u_0=0.1,0.2,...,0.9$. Inset (a)
gives absolute error of this expression with respect to a straight
line approximation (for the same set of $u_0$). Inset (b) shows
the relative error between the result of numerical integration in
the lhs of Eq.~(\ref{eq:app:sigma:int}) and ${\frak W}u_0^{3/2}$
as a function of $n_B$ (solid line). The dashed line is a linear
fit of the form $0.035n_B$.}\label{figB.eps}
\end{figure}
\section{Surface Tension Near the Transition Curve}\label{app:sigma}

The surface tension has been defined as
%/////////////////////////////////////////////////////
\begin{equation}\label{eq:app:sigma:def}
\sigma  = \sqrt{\frac{\hbar^2g_0^2{\frak E}}{2m_Ba_{BB}^3}}
\frac{u_0^{1/2}}{2A}\int\limits_0^1 {d\sqrt z } \sqrt {\varepsilon
(u_0 z) +1 }
\end{equation}
%/////////////////////////////////////////////////////
where we used the notations of Appendix~\ref{app:shape}. Utilizing
the approximation (\ref{eq:shape:approxE}) at the i-ii transiiton
curve, i.e. setting $u_1=0$, we obtain
%/////////////////////////////////////////////////////
\begin{equation}\label{eq:app:sigma:int}
\int\limits_0^1 {d\sqrt z} \sqrt{\varepsilon(u_0 z)+1} = {\frak
W}u_0^{3/2} \left[1+\mathcal{O}(n_B/100)\right]
\end{equation}
%/////////////////////////////////////////////////////
The left-hand side can be integrated numerically for various $u_0$
to verify the result, see Fig.~\ref{figB.eps}. Significant
relative error of Eq.~(\ref{eq:shape:approxE}) at $u\to u_0$ is
suppressed in the above integral and the approximation
(\ref{eq:app:sigma:int}) is accurate along the entire phase
separation curve. Using the value of ${\frak M}$ found earlier
(see Appendix~\ref{app:shape}) we finally obtain
%/////////////////////////////////////////////////////
\begin{equation}\label{eq:app:sigma:final}
\sigma = 0.304\frac{\sqrt\pi\hbar^2}{2m_Ba_{BB}^4}g_0^3n_B^2
\left[1+\mathcal{O}(1/100)\right]
\end{equation}
%/////////////////////////////////////////////////////

For large $n_B$, closer to ``tricritical" point, the effective
surface tension will deviate from the one obtained above due to
additional gradient terms in Eq.~(\ref{eq:equil:S}). This is due
to renormalization of the boson kinetic energy which is ignored by
the Thomas-Fermi approximation. This correction is negligible for
the parameters of interest, as explained in Sec.~\ref{sec:equil}.

%/////////////////////////////////////////////////////
%////////////////////////////////////////////////////////////
%////////////////////////////////////////////////////////////
%////////////////////////////////////////////////////////////


\begin{thebibliography}{99}{\frenchspacing

\bibitem{UedaLeggett} H. T. C. Stoof, J. Stat. Phys. \textbf{87}, 1353 (1997); M. Ueda and A. J.
Leggett, Phys. Rev. Lett. \textbf{80}, 1576 (1998).

\bibitem{Modugno} G. Roati, M. Zaccanti, C. D'Errico, J. Catani, M. Modugno, A.
Simoni, M. Inguscio, and G. Modugno, Phys. Rev. Lett. 99, 010403
(2007).

\bibitem{LifshitzKagan} I. M. Lifshitz and Yu. Kagan, Zh. Eksp. Teor. Fiz. \textbf{62}, 385 (1972) [Sov. Phys. JETP
\textbf{35}, 206 (1972)].

\bibitem{LifshitzKhokhlov} I. M. Lifshitz, V. N. Polesskii, and V. A. Khokhlov,
Zh. Eksp. Teor. Fiz. \textbf{74}, 268 (1978) [Sov. Phys. JETP
\textbf{47}, 137 (1978)].

\bibitem{Burmistrov} S. N. Burmistrov, L. B. Dubovskii, and V. L.
Tsymbalenko, J. of Low Temp. Phys. \textbf{90}, 363 (1993).

\bibitem{Edwards} R. De Bruyn Ouboter, et al., Physica 26, 853 (1960); D. O. Edwards
and J. G. Daunt, Phys. Rev. 124, 640 (1961); D. O. Edwards et al.,
Phys. Rev. Lett. 15, 773 (1965).

\bibitem{HeExperiment} V. A. Mikheev et al., Phys. Low Temp. (U.S.S.R.) \textbf{17}, 444 (1991);
T. Satoh et al., Phys. Rev. Lett. \textbf{69}, 335 (1992).

\bibitem{SELF} D. Solenov and D. Mozyrsky, Phys. Rev. Lett. 100, 150402 (2008).


\bibitem{BosCool} A. G. Truscott et al., Science \textbf{291}, 2570 (2001); M. W. Zwierlein
et al., Phys. Rev. Lett. \textbf{92}, 120403 (2004).

\bibitem{Ketterle} Y. Shin et al., Phys. Rev. Lett. \textbf{97}, 030401 (2006).

\bibitem{bosons} Unlike in fermion-boson
mixtures, phase separation transition in binary boson-boson
systems corresponds to a continuous second  order-like transition;
see Ref.~\onlinecite{Timmermans}

\bibitem{Timmermans} E. Timmermans, Phys. Rev. Lett. \textbf{81}, 5718
(1998).

\bibitem{Mahan} G. D. Mahan, \emph{Many-Particle Physics\/} (Kluwer Academic, New York, 2000).

\bibitem{Viverit} L. Viverit, C. J. Pethick, and H. Smith, Phys. Rev. A \textbf{61},
053605 (2000).

\bibitem{Mozyrsky} D. Mozyrsky, I. Martin, and E. Timmermans, Phys. Rev. A
\textbf{76}, 051601(R) (2007).

\bibitem{Kleinert} H. Kleinert, \emph{Path Integrals in Quantum Mechanics, Statistics,
Polymer Physics, and Financial Markets\/} (World Scientific
Publishing, 2004).

\bibitem{ImambekovDemler} A. Imambekov, C. J. Bolech, M. Lukin, and E.
Demler, Phys. Rev. A 74, 053626 (2006).

\bibitem{LifshitzPitaevskii} E. M. Lifshitz and L. P. Pitaevskii, Physical Kinetics, (Pergamon
Press, 1981).

\bibitem{Coleman} S. Coleman, \textit{Aspects of Symmetry} (Cambridge University Press, 1985).

\bibitem{CaldeiraLeggett} A. O. Caldeira and A. J. Leggett, Phys. Rev. Lett. \textbf{46}, 211
(1981).

\bibitem{DSVB} D. Solenov and V. A. Burdov, Phys. Rev. B 72, 085347, (2005);

\bibitem{DMmosfet} D. Mozyrsky, I. Martin, A. Shnirman, and M. B. Hastings,
cond-mat/0312503 at www.arXiv.org.

\bibitem{DSmosfet} D. Solenov, Phys. Rev. B 76, 115309 (2007).

\bibitem{VBDS} V. A. Burdov and D. Solenov, Int. J. of Nanoscience 6, 389
(2007).

\bibitem{Ls-note} Note that $r/x$ coincides with $l_s$ found earlier.

\bibitem{Gorokhov} D. A. Gorokhov and G. Blatter, Phys. Rev. B \textbf{56}, 3130 (1997).

\bibitem{Kamenev}  A. Kamenev, cond-mat/0412296 at www.arXiv.org

 }\end{thebibliography}
\end{document}